\documentclass[onecolumn,12pt]{IEEEtran}

\usepackage{times}
\usepackage{amsmath,amssymb,mathrsfs}
\usepackage{graphicx}
\usepackage{color}
\usepackage{subfigure}
\usepackage{cite}
\usepackage{multirow,tabularx}
\usepackage[ruled,lined]{algorithm2e}

\usepackage{wasysym}
\usepackage{stackrel}

\newcommand{\qed}{\nobreak \ifvmode \relax \else
      \ifdim\lastskip<1.5em \hskip-\lastskip
      \hskip1.5em plus0em minus0.5em \fi \nobreak
      \vrule height0.75em width0.5em depth0.25em\fi}

\newcommand{\beq}{\begin{equation}}
\newcommand{\eeq}{\end{equation}}

\newcommand{\bdisp}{\begin{displaymath}}
\newcommand{\edisp}{\end{displaymath}}

\newcommand{\beqarr}{\begin{eqnarray}}
\newcommand{\eeqarr}{\end{eqnarray}}

\newcommand{\bmlt}{\begin{multline}}
\newcommand{\emlt}{\end{multline}}

\newcommand{\beqarrn}{\begin{eqnarray*}}
\newcommand{\eeqarrn}{\end{eqnarray*}}

\newcommand{\benum}{\begin{enumerate}}
\newcommand{\eenum}{\end{enumerate}}

\newcommand{\bit}{\begin{itemize}}
\newcommand{\eit}{\end{itemize}}

\newcommand{\bc}{\begin{center}}
\newcommand{\ec}{\end{center}}

\newcommand{\bdes}{\begin{description}}
\newcommand{\edes}{\end{description}}

\newcommand{\bfig}{\begin{figure}}
\newcommand{\efig}{\end{figure}}

\newcommand{\bemq}{\begin{quote} \begin{em}}
\newcommand{\eemq}{\end{em} \end{quote}}

\newcommand{\bmp}{\begin{minipage}}
\newcommand{\emp}{\end{minipage}}

\newcommand{\eqn}[1]{(\ref{#1})}

\newcommand{\etal}{{\it et al.}}

\newcommand{\bsp}{\begin{slide*}}
\newcommand{\esp}{\end{slide*}}
\newcommand{\bsl}{\begin{slide}}
\newcommand{\esl}{\end{slide}}

\newcommand{\barb}[1]{\Bar{ #1 }}

\newcommand{\virag}[1]{}

\IEEEoverridecommandlockouts

\newcommand{\abs}[1]{\left\lvert{#1}\right\lvert}

\newcommand{\fnet}[1]{#1}

\newcommand{\net}{\mathcal{N}}
\newcommand{\netnodes}{V}
\newcommand{\netedges}{E}

\newcommand{\netsources}{S}

\newcommand{\neighbors}[1]{N (#1)}

\newcommand{\compgraph}{\mathcal{G}}
\newcommand{\compnodes}{\Omega}
\newcommand{\compedges}{\Gamma}

\newcommand{\numS}{\kappa}
\newcommand{\numT}{\gamma}
\newcommand{\preedges}[1]{\Phi_{\uparrow}(#1)}
\newcommand{\sucedges}[1]{\Phi_{\downarrow}(#1)}

\newcommand{\head}[1]{head(#1)}
\newcommand{\tail}[1]{tail(#1)}
\newcommand{\Ell}{L}
\newcommand{\br}{{\bf r}}

\newcommand{\Func}{\Theta}
\newcommand{\cA}{\mathcal{A}}

\newcommand{\embedding}{B}
\newcommand{\setofembeddings}{\mathcal{B}}
\newcommand{\setofpaths}{\mathcal{P}}
\newcommand{\pathstart}[1]{\text{start}(#1)}
\newcommand{\first}[1]{\text{start}(#1)}

\newcommand{\pathend}[1]{\text{end}(#1)}

\newcommand{\weight}[2]{w_{#1}(#2)}
\newcommand{\minweight}[1]{\alpha_{\Ell}}
\newcommand{\numgraphs}{\nu}
\newcommand{\by}{{\bf y}}

\newcommand{\selfandneighbors}[1]{N' (#1)}

\newcommand{\defn}{\stackrel{\triangle}{=}}

\newcommand{\precision}[1]{b(#1)}
\newcommand{\energy}{E}

\begin{document}

\title{Network Flows for Functions}

\author{
\authorblockN{Virag Shah \hspace{5mm} Bikash Kumar Dey \hspace{5mm} D. Manjunath}  \\
\authorblockA{Department of Electrical Engineering  \\
Indian Institute of Technology Bombay \\
Mumbai, India, 400 076\\
virag4u@gmail.com,\{bikash,dmanju\}@ee.iitb.ac.in}
}

 \maketitle

\begin{abstract}

We consider in-network computation of an arbitrary function over an
arbitrary communication network. A network with
capacity constraints on the links is given. Some nodes in
the network generate data, e.g., like sensor nodes in a sensor
network. An arbitrary function of this distributed data is to be
obtained at a terminal node. The structure of the function is
described by a given computation schema, which in turn is represented
by a directed tree. We design computing and communicating 
schemes to obtain the function at the terminal at the
maximum rate. For this, we formulate linear programs to determine
network flows that maximize the computation rate. We then develop
fast combinatorial primal-dual algorithm to obtain $\epsilon$-approximate
solutions to these linear programs. We then briefly describe extensions
of our techniques to the cases of multiple terminals wanting different
functions, multiple computation schemas for a function, computation with
a given desired precision, and to networks with energy constraints
at nodes.

\end{abstract}

\IEEEpeerreviewmaketitle

\section{Introduction}
\label{sec:intro} 

Motivated by sensor network applications, there has been significant
interest in computing functions of distributed data inside the
network. A typical scenario that is considered is as follows. Sensor
nodes, distributed in a sensor field, can make measurements of their
environment, perform reasonable amounts of computation and also
communicate with other nodes. The interest of the sensor network is
not so much in the measurement values made by the sensors but of some
function of these variables, say $\Func.$ Since the nodes in the
network can perform computation, they could participate in the
computation of $\Func.$ Thus the interest is in distributed
computation of a function of distributed data. This has also been
called `in-network function computation.'  In this setting, it is
typically assumed that the variables form a time sequence and that
they can be generated at any rate; equivalently, an infinite sequence
is readily available. Thus, in this setting it is natural to want to
compute $\Func$ at the best rate possible.  
In this paper, we
introduce novel network flow techniques to design a computation and
communication scheme that maximizes the rate at which $\Func$ is
computed. Though network flow techniques have been used widely
to study multiple unicast~\cite{leighton,shahrokhi,Ahuja93,garg} problems, our work develops such techniques
for the first time for function computation.

Early work on in-network computation was on the asymptotic analysis of
the number of transmissions needed to compute specific functions in
noisy broadcast networks. e.g.,
\cite{Gallager88,Kushilevitz98,Feige00}. In recent works, it is assumed
that the node locations are from a realization of a suitable random
point process, hence the resulting communication graph of the network
is a random graph, e.g., \cite{Giridhar05,Ying06,Kanoria07,Kamath08}.
In this setting a probabilistic characterization of the asymptotic (in
the number of nodes) computation rate for different classes of
functions, such as `type-threshold
functions' and 'type sensitive functions' \cite{Giridhar05},
have been obtained.

Another class of work considers simplistic networks with small
number of correlated sources~\cite{korner1,han1,orlitsky1,feng1}. Much of this work takes the information
theoretic perspective in which the objective is to find encoding
rate regions for reliably communicating the desired function.
This class of work allows block coding to achieve better rates.
There has been some recent work in the network coding literature 
on distributed function computation~\cite{RaiD:09c,appuswamy3,langberg3}.
They consider larger and more complex 
networks with independent sources. However, designing optimal coding
schemes and finding capacity is a difficult problem except for
very special functions or networks~\cite{RaiD:09c,appuswamy3}.

 In this paper we make a significant departure from
the above. We consider arbitrary functions of
the distributed data for which a computation schema is described by a
directed tree. A computation schema defines a sequence of operations
to compute the function. An arbitrary communication network over which $\Func$ is to be
computed is assumed given. Our techniques work for networks with
both directed as well as undirected links with capacity constraints,
though we present our results only for networks with undirected links.
There are some similarities of our work with that of graph
embedding.  e.g., \cite{Leighton92,Wohlmuth98,Heun02} but there are
significant differences in the modeling assumptions and in the
embedding objectives. Such work typically assume the target network to
be a `regular network' like a hypercube or a mesh and all link
capacities are assumed equal. The embedding objective is to minimize
the parameters like `dialation.'

\subsection{An Example and Motivation}
\begin{figure*}[!t]
\centering
\subfigure[]{
\includegraphics[width=.2\textwidth,height=0.25\textwidth]{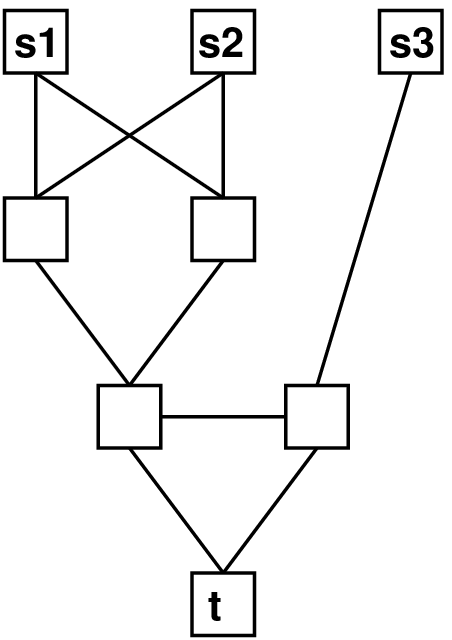}
\label{fig:network}
}
\subfigure[]{
\includegraphics[width=.2\textwidth,height=0.25\textwidth]{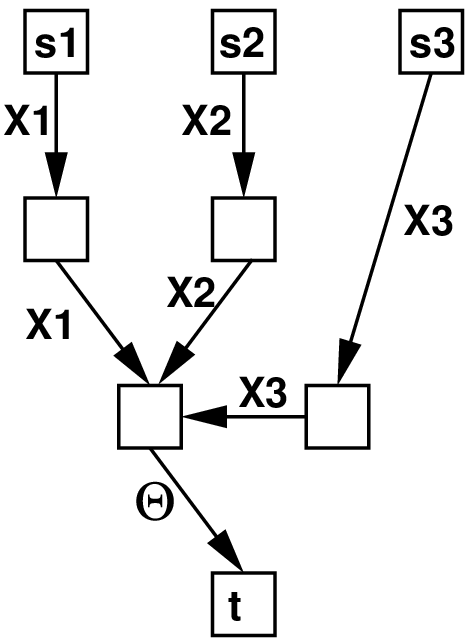}
\label{fig:embed1}
}
\subfigure[]{
\includegraphics[width=.2\textwidth,height=0.25\textwidth]{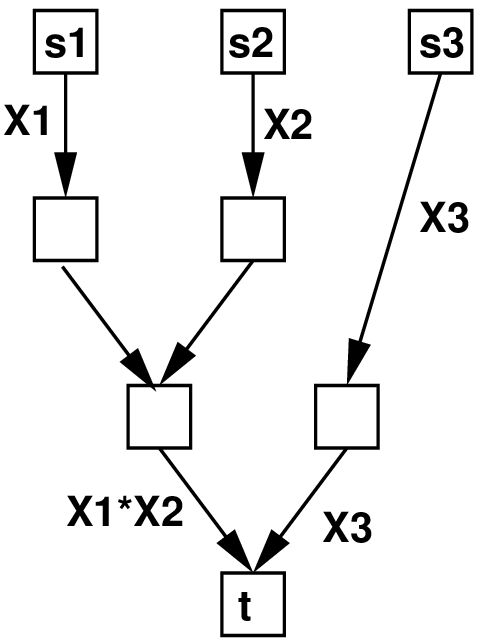}
\label{fig:embed2}
}
\subfigure[]{
\includegraphics[width=.2\textwidth,height=0.25\textwidth]{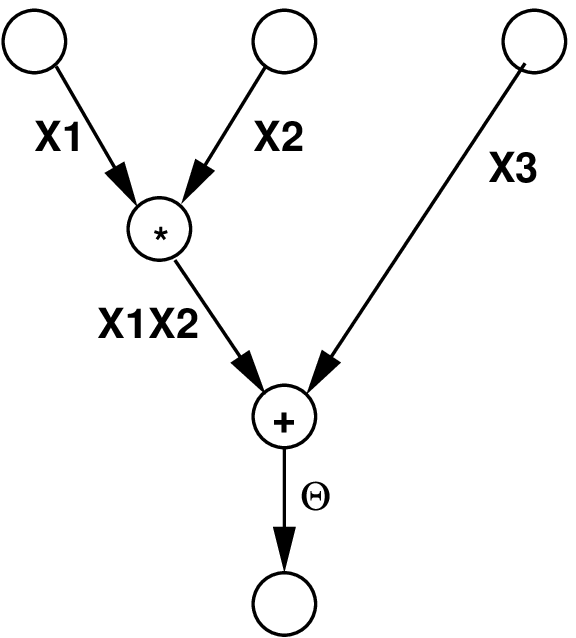}
\label{fig:comptree}
}
  \caption{Computing $\Func = X_1 X_2 + X_3$ over a network. (a) A
    network to compute $\Func = X_1 X_2 + X_3.$ (b) A possible
    embedding that computes at $\Func$ at unit rate. (c) An
    alternative embedding. (d) A schema to compute $\Func.$}
\label{fig:example1}
\end{figure*}

Let us consider the function $\Func (X_1, X_2, X_3) = X_1X_2 + X_3$ of
three variables generated at three sources
$s_1, s_2,$ and $s_3$ respectively.
A terminal node $t$
is required to obtain the function $\Func (X_1, X_2,
X_3)$. We assume that all the three data symbols are from the same
alphabet $\cA$. The computation of the function can be broken into
two parts, namely, first computing $X_1X_2$, and then adding
$X_3$. These two operations can be done at different nodes in the
network in the above order. This decomposition of the computation can
be represented by the graph shown in Fig.~\ref{fig:comptree}.
Such a graphical representation of the computation will henceforth be
called a computation tree.
Each edge represents a {\it unique} function of the source symbols.

Now consider computing $\Func (X_1, X_2, X_3)$ in the network shown in
Fig.~\ref{fig:network} where each edge has unit capacity. There are multiple
ways of receiving this function at the terminal $t$ depending on
what computations are done at what nodes and along what
paths the data flows. Two such ways of computing this function
are shown in Figs.~\ref{fig:embed1} and \ref{fig:embed2}.
These are called `embeddings', defined formally in Sec.~\ref{sec:model}.
It is clear that intelligent time-sharing
between these various embeddings may give higher number of computation
per use of the network on average than using only one such
embedding. This raises the natural question: what is the maximum rate
of computing that can be achieved on a given network and how to
achieve it?

\subsection{Organization and Summary of Contributions}
We begin by describing the model in detail in the next section.
Section~\ref{sec:tree} presents the main contributions of this
paper. Here we formulate a linear program, \emph{Embedding-Edge-LP,}
that optimally allocates flows on the embeddings.
We then present another LP,
\emph{Node-Arc-LP,}, based on a flow conservation law. This
LP can be solved in polynomial time. We then describe an
algorithm, \emph{Algorithm~\ref{schedulingalgo},} that converts the flow rates
obtained from \emph{Node-Arc-LP} into a flow allocation on the embeddings.
We then present a fast primal-dual algorithm which finds a
solution to achieve at least $(1-\epsilon)$ fraction of the
optimal rate. We call such a solution an $\epsilon$-approximate solution.
This algorithm uses an oracle subroutine which finds
a minimum cost embedding of the computation tree in the network.
We provide an efficient algorithm, {\em OptimalEmbedding($\Ell$)},
to obtain the same.
 This algorithm is also of independent interest.
Four interesting extensions of our results are presented in
Sec.~\ref{sec:extensions}. First, we allow multiple computing
schema for computing the same function. Then we consider
multiple terminals computing distinct functions of disjoint sets
of sources. For this problem, we modify our techniques to
maximize the
weighted sum-rate of computations, and also to maximize the rate-tuple
in a given direction. In the third extension, we consider the problem
of computing a function with a desired precision which is achieved
by allowing possibly different precision for each type of data.
In the fourth extension, we consider a network with energy-constrained
nodes, and assume that each type of data, i.e., each edge of the
computation tree, requires some fixed but different amount of energy
to compute/generate, transmit, and receive.

\section{The model and the notation }
\label{sec:model}

The communication network is an undirected, simple, connected graph
$\net =(\netnodes,\netedges)$ where $\netnodes$ is a set of
$n$ nodes and $\netedges$ is a set of $m$ undirected edges. Each edge
$uv \in \netedges$ represents a half duplex link with a total
non-negative capacity $c(uv).$
In the network, $\netsources = \{ \fnet{s_1, s_2, \ldots, s_\numS} \}
\subset \netnodes$ is the set of $\numS$ source nodes.  Source $s_i$
has an infinite sequence of data values $\{X_i(k)\}_{k \geq 0}$ where
$X_i(k)$ belongs to a finite alphabet $\cA.$ The link capacities
are expressed in $|\cA|$-ary unit. $X_i$ is used to denote a
representative element of the sequence.  Let $X \defn [X_1, \ldots
X_{\numS}].$ Without loss of generality, we assume that each source
node in the network generates exactly one data sequence. If a source
node generates two or more data sequences then this can be represented
by multiple source nodes connected by infinite capacity links.  We
also assume that there is only one terminal node. 

A given function $\Func: {\cA}^{\numS} \to \cA$ of $X$ 
needs to be obtained at the terminal
node $t$ for each $k$ at the best possible rate.  A computation
schema for $\Func$ is given and represented by a directed tree
$\compgraph = (\compnodes, \compedges)$ where $\compnodes$ is the set
of nodes and $\compedges$ is the set of edges. The elements of
$\compnodes$ are labelled $\mu_1, \mu_2, \ldots, \mu_{|\compnodes|}$
where $\mu_1, \mu_2, \ldots, \mu_\numS$ are the source nodes,
$\mu_{|\compnodes|} $ is the terminal node that obtains $\Func$
and the rest are computing nodes that compute different functions of
$X.$ Further, the nodes in $\compnodes$ are labelled according to a
topological order such that for $i > j$ there is no directed path in
$\compgraph$ from $\mu_i$ to $\mu_j.$ The source nodes have in-degree
 zero and out-degree  one and the terminal node has in-degree one and
out-degree zero.  All other nodes have in-degree greater than one and 
out-degree exactly one. Similarly, the elements of $\compedges$ are labelled
$\theta_1, \theta_2, \ldots, \theta_{|\compedges |}$ with $\theta_1,
\theta_2, \ldots, \theta_\numS$ being the outgoing edges from $\mu_1,
\mu_2, \ldots, \mu_\numS$ respectively, and $\theta_{|\compedges|}$ being
the incoming edge into $\mu_{|\compnodes|}$. The remaining edges are labeled according to a topological order, i.e., for any $i<j$, there is no path from the head node of $j$ to the tail node of $i$. The nodes and edges of $\compgraph$ can be labeled as above in $O(|\compedges|)= O(\numS)$ time.

For any edge $\theta \in \compedges$, let $\tail{\theta}$ and
$\head{\theta}$ represent, respectively, the tail and the head nodes of
the edge $\theta.$ Let $\preedges{\theta}$ and $\sucedges{\theta}$
denote, respectively, the predecessors and the successors of $\theta,$
i.e.,
\begin{eqnarray*}
  \preedges{\theta} & \defn & \{\eta \in \compedges | \head{\eta}
  = \tail{\theta} \}
  \mbox{ and } \\
  \sucedges{\theta} & \defn & \{\eta \in \compedges | \tail{\eta}
  = \head{\theta} \}.
\end{eqnarray*}
Each edge $\theta$ of $\compgraph$ represents a distinct function of $X$ that
can be computed from the functions corresponding to the edges
in $\preedges{\theta}$. Further, each function takes values from
the same alphabet $\cA$. (We remark here that this is not
unreasonable even when all the computations are over real numbers
because computations are performed using a fixed precision.)

Let $\neighbors{v} \defn \{u \in \netnodes | uv \in \netedges \}$ 
denote the set of neighbors of a node $v \in \netnodes.$ We also
denote the set of neighbors and itself by $\selfandneighbors{v}
= \neighbors{v}\cup \{v\}$. A sequence of nodes $v_1,v_2,\cdots,
v_l$, $l\geq 1$, is called a path if $v_iv_{i+1}\in E$ for $i=1,2,\ldots, l-1$.
The set of all paths in $\net$ is denoted by $\setofpaths$.
With abuse of notation, for such a path $P$, we will say $v_i \in P$ and
also $v_iv_{i+1}\in P$. The nodes $v_1$ and $v_l$ are called
respectively the start node and the end node of $P$, and are denoted
as $\pathstart{P}$ and $\pathend{P}$.

As discussed in Sec.~\ref{sec:intro}, a function with a given computation
tree can be computed along any ``embedding'' of the tree in the
network as shown in Fig.~\ref{fig:example1}. We are now ready to formally define an embedding of
a computation tree.

{\bf Definition:} An embedding is a mapping $\embedding : \compedges \to \setofpaths$ such that

\begin{enumerate}
\item $\pathstart{B(\theta_l)} = s_l$ for $l=1,2,\ldots,\numS$
\item $\pathend{B(\eta)} = \pathstart{B(\theta)}$ if $\eta \in \preedges{\theta}$
\item $\pathend{B(\theta_{|\compedges|})} = t$.
\end{enumerate}

We denote the set of embeddings of $\compgraph$ in $\net$ by $\setofembeddings$.
Our aim is to determine the flows on these embeddings so as to
maximize the total flow.
An edge in the network may carry different functions of the source
data in an embedding. We thus define the number of times an edge $e\in \netedges$
is used in an embedding $\embedding$ as  $r_{\embedding}(e) = |\{\theta \in \compedges | e \mbox{ is a part of }
\embedding(\theta )\}|$.  Note that $|r_{\embedding}(e)| \leq |\compedges|$
for any edge, and $r_{\embedding}(e) = 0$ for an edge $e$ which
is not used by the embedding $\embedding$. Further, an edge may also
be used to carry flows on different embeddings.
Therefore in an assignment of flows on different embeddings, i.e.,
in a particular timesharing scheme, the edge may carry multiple
types of data (i.e., different functions of $X$) of
different amounts.

\section{Linear programs and algorithms}
\label{sec:tree}

\newcommand{\alwaysremove}[1]{}

In this section, we present our main contributions. 
\begin{itemize}
\item In Section~\ref{sec:embed}, we give a basic linear program, the
  {\em Embedding-Edge LP}, which characterizes our problem.
\item In Section~\ref{sec:node-arc}, we give an alternate LP, the
  {\em Node-Arc LP}, that can be solved in polynomial time. We then present
  an algorithm which obtains a solution of the {\em Embedding-Edge LP} with
  the same rate from a solution of the {\em Node-Arc LP}.
\item Drawing parallels from multi-commodity flow techniques, we give,
  in Section~\ref{primal_dual_algo}, the dual of our {\em Embedding-Edge LP}
  and present a fast primal-dual algorithm to compute an
  $\epsilon$-approximate solution. This algorithm needs a subroutine
  which finds a `minimum weight embedding' of the computation tree in
  the network for given edge-weights. We present an efficient exact
  algorithm for this purpose. This algorithm is of independent
  interest, for instance, for computing functions over a network with
  power limited, but with infinite bandwidth, links.
\end{itemize}

\newcommand{\barr}{\begin{array}}
\newcommand{\earr}{\end{array}}

Note that, if $\pathstart{\embedding({\theta_i})} =
\pathend{\embedding({\theta_i})}$, i.e., if $\embedding({\theta_i})$
consists of a single node, then in that embedding the data $\theta_i$
is generated as well as used (i.e., not forwarded to another node) in
that node.

\subsection{The Embedding-Edge LP}
\label{sec:embed}

As discussed in Sec.~\ref{sec:intro} and Sec.~\ref{sec:model}, the function for a particular
sample of the data can be computed over the network using any
embedding of the computation tree in the network. Let $\setofembeddings$
be the set of all embeddings of $\compgraph$ in $\net$. For any
embedding $\embedding \in \setofembeddings$, let $x(B)$ denote
the average number of function symbols computed using the embedding
$B$ per use of the network.
We present below a linear program to maximize the computation rate $\lambda
= \sum_{\embedding \in \setofembeddings} x(\embedding)$.
Recall that $r_{\embedding}(e)$ represents the number of times the edge $e$ is used
in the embedding $\embedding$.

\centerline{\rule{\columnwidth}{0.75pt}}
{\em Embedding-Edge LP:} Maximize $\lambda = \sum_{\embedding \in \setofembeddings} x(\embedding)$ subject to

1. Capacity constraints
\begin{equation}
  \sum_{\embedding \in \setofembeddings }
  r_{\embedding}(e)x(\embedding) \le c(e), \; \forall e \in \netedges
\end{equation}

2. Non-negativity constraints
\begin{equation}
  x(\embedding) \ge 0, \; \forall \embedding
\end{equation}
\centerline{\rule{\columnwidth}{0.75pt}}

This LP finds an optimal fractional packing of the embeddings
of $\compgraph$ into $\net$. Similar formulations have been
considered widely in literature in the context of multi-commodity
flow~\cite{garg,karakostas} and other packing problems~\cite{garg}.

In multi-commodity flow problems, a solution of the so called
{\it Path-edge LP} readily gives a way of achieving the corresponding
rates. However, since in our problem, the data is to be mixed
according to different embeddings for different realizations of data, one
needs to carefully device a protocol to schedule the computation
and communication at the nodes and edges in such a way that
data from different realizations are not mixed. Such a protocol
is presented in the appendix.

\subsection{The Node-Arc LP}
\label{sec:node-arc}

\begin{figure}
  \centering{\includegraphics[width=4in]{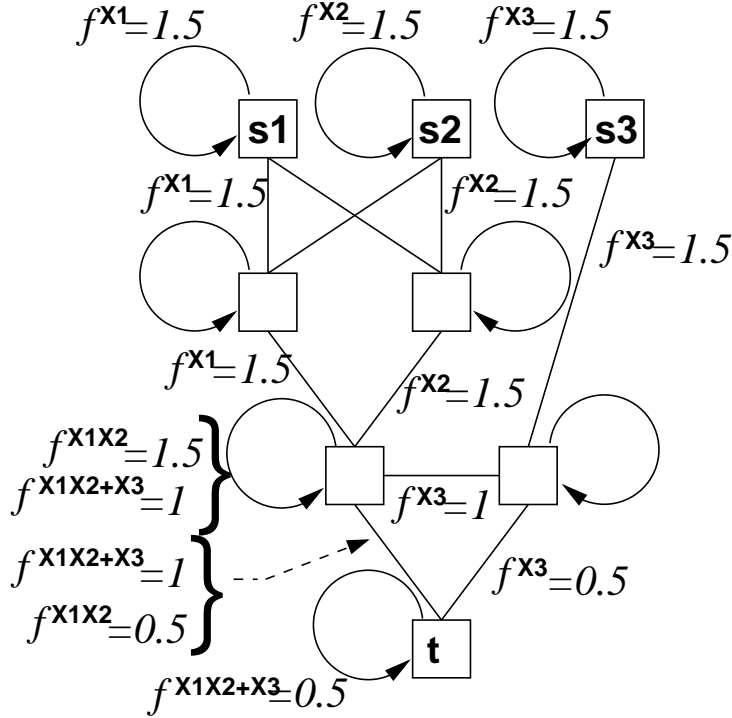}}
  \caption{The aggregate edge-flow values for a flow of 0.5 on the embedding
in Fig.~\ref{fig:embed2} and a flow of 1 on the embedding in Fig.~\ref{fig:embed1}.}  \label{fig:nodearc}
\end{figure}

Note that the cardinality of $\setofembeddings$ can be exponential
in $|\netnodes |$. Hence the complexity of the {\em Embedding-Edge LP} is exponential
in the network
parameters if any other structure of the problem is not used.
In the multi-commodity flow literature, another LP formulation,
called the {\em Node-Arc LP}, based on the flow conservation principle
is well-known which can be solved in polynomial time.
In the following, we formulate a node conservation based LP
for our problem.
For this LP, we assume that each node in the network has a virtual
self-loop of infinite capacity. The data flowing in the self-loop
represents the data generated at that node. This may be the source
data generated at the sources or the intermediate or final values
computed at the node. For example, if a node computes $X_1X_2$ from $X_1$ and
$X_2$ it receives, and then computes $X_1X_2+X_3$ by using the computed
$X_1X_2$ and received $X_3$, then both $X_1X_2$ and $X_1X_2+X_3$ will be assumed
to be flowing in its self-loop. Example of the flows on the edges
and the self-loops corresponding to a particular flow assignment on
two embeddings is shown in Fig.~\ref{fig:nodearc}.

The variables in our {\em Node-Arc LP} are
\begin{eqnarray}
& & \left\{f^\theta_{uv},f^\theta_{vu}|uv \in \netedges, \theta \in \compedges \right\}\cup \left\{f^\theta_{uu}|u \in \netnodes, \theta \in \compedges \right\}\cup\{\lambda\}. \nonumber
\end{eqnarray}
where, $f^\theta_{uv}$ represents the flow of type $\theta \in \compedges$ flowing through the edge $ uv \in 
\netedges$ from $u$ to $v$, $f^\theta_{uu}$ denotes the flow of type $\theta$ flowing in the
self-loop at $u$ and $\lambda$ represents the total rate of the function computation.

The linear program consists
of capacity constraint on the edges of $\net$, a flow-conservation rule on the nodes of $\net$, and non-negativity
constraint on the flows $f^\theta_{uv}$.
The flow conservation rule is based on the fact that an intermediate node
in $\net$ can, apart from forwarding 
the flows it receives, generate a flow of type $\theta$ on its self-loop
by terminating the same amount of incoming flows of type
$\eta \in \preedges{\theta}$.
Each source node $s_l$, in addition, generates
$\lambda$ amount of flow of type $\theta_l$.
Similarly, the  terminal node $t$ terminates
$\lambda$ amount of flow of type $\theta_{|\compedges|}$.
The {\em Node-Arc LP} is as follows. Recall that $\selfandneighbors{v}$
denotes the set of the neighbors of $v$ and itself.

\centerline{\rule{\columnwidth}{0.75pt}}

{\em Node-Arc LP:} Maximize $\lambda$ subject to following
constraints  any node $v\in \netnodes$

1. Functional conservation of flows: 
\begin{eqnarray}
 &&f^{\eta}_{vv}  
  + \sum_{u \in \neighbors{v}}  f^{\theta}_{vu}  - \sum_{u \in \selfandneighbors{v}}   f^{\theta}_{uv}  = 0
  , 
\hspace*{2mm} \forall \theta \in \compedges \setminus \{\theta_{|\compedges|}\} \; \text{and} \;
  \forall \eta \in \sucedges{\theta}.
\label{eq:general_conservation}
\end{eqnarray}

2. Conservation and termination of $\theta_{|\compedges|}$: 
\begin{equation}\label{eq:flow_termination}
\sum_{u \in \neighbors{v}}  f^{\theta_{|\compedges|}}_{vu}  - \sum_{u \in \selfandneighbors{v}}   f^{\theta_{|\compedges|}}_{uv}  = 
\begin{cases}
- \lambda & v=t \\
0. & \text{otherwise}
\end{cases}
\end{equation}

3. Generation of $\theta_l \; \forall l \in \{1,2,\ldots, \numS\}$:
\begin{equation}\label{eq:source_generation}
 f^{\theta_l}_{v v} = 
\begin{cases}
\lambda & v=s_l \\
0. & \text{otherwise}
\end{cases}
\end{equation}

4. Capacity constraints

\begin{equation}
  \sum_{\theta \in \compedges} \left( f^{\theta}_{uv} + f^{\theta}_{vu}
\right) \le c(uv) , \; \forall uv \in \netedges .
\end{equation}

5. Non-negativity constraints 

\begin{eqnarray}
&& f^{\theta}_{uv} \ge 0 , \forall uv \in \netedges \;  \text{and} \; \forall \theta \in \compedges  \\
&& f^{\theta}_{uu} \ge 0 , \forall u \in \netnodes \;  \text{and} \; \forall \theta \in \compedges  \\
&& \lambda \ge 0 .
\end{eqnarray}

\centerline{\rule{\columnwidth}{0.75pt}}

This LP has $O(\numS m)$ number of variables, $O(\numS m)$ number
of non-negativity constraints (one for each variable), and $O(\numS n + m)$
number of other constraints. Hence it can be solved in polynomial
time. 

The above LP gives a set of flow values on each link. Now we briefly describe
and present an algorithm,
Algorithm~\ref{schedulingalgo}, which, from any 
feasible solution of this LP, obtains a corresponding
feasible solution for the {\em Embedding-Edge LP} that achieves the same $\lambda$.

Each iteration of the {\em while} loop finds an embedding with a non-zero
flow and removes the corresponding edge-flows to
obtain another feasible solution with a reduced rate.
This continues until $\lambda$ amount of flow has been extracted.
The $i$-th iteration of the {\em for} loop finds the mapping of $\theta_i$
in the embedding. While exploring the nodes to find the mapping of
$\theta_i$, it checks for the presence of a cycle of flow of type
$\theta_i$. It removes such a cycle if detected.

\begin{algorithm}\label{schedulingalgo} \caption{Finding  equivalent solution of the {\em Embedding-Edge LP} from a feasible solution of the {\em Node-Arc LP}. } 
  \SetKwData{Left}{left}\SetKwData{This}{this}\SetKwData{Up}{up}
  \SetKwFunction{Union}{Union}\SetKwFunction{FindCompress}{FindCompress}
  \SetKwInOut{Input}{input}\SetKwInOut{Output}{output}
  
  \Input{Network graph $\net =(\netnodes,\netedges)$, capacities $c(e)$, set of source
    nodes $\netsources$, terminal node $t$, computation tree $\compgraph = (\compnodes, \compedges )$,
    and a feasible solution  to its {\em Node-Arc LP} that consists of the values of $\lambda$, $f^{\theta}_{uv}$ $\forall \theta \in \compedges$, $\forall uv \in \netedges$, and $f^{\theta}_{uu}$ $\forall \theta \in \compedges,\,\,\forall u \in \netnodes$. }

\Output{Solution $\{x(\embedding )|\embedding \in \setofembeddings \}$ to
the {\em Embedding-Edge LP} with $\sum_{\embedding \in \setofembeddings}x(\embedding ) = \lambda$.} 
  
\BlankLine Initialize $x(\embedding) := 0, \embedding (\theta_i) = \emptyset
   \text{ (the null sequence)},
   \forall \embedding \in \mathcal{\setofembeddings} \text{ and }
   \forall \theta_i \in  \compedges, \lambda'=0$

\While{$\lambda' \neq \lambda$ }{ 
   $z(t)$ := $\lambda$ \; $\embedding(\theta_{|\compedges|}):=t$ \;
   \For{$i$ := $|\compedges|$ \KwTo $1$}{
     $v$ := $\embedding(\theta_i)$ \tcp*{valid, as $B(\theta_i)$ has of only one node at this step}
     $u$ := an element in $\selfandneighbors{v}$ such that $f^{\theta_i}_{uv} >0$ \;
     \If{ $u \neq v$ and $u \in \embedding (\theta_i )$}{
      \tcp{A cycle of redundant flow found: remove the flow from all the edges
           in the cycle}
      Let $P$ be the path in $\embedding (\theta_i )$ upto the first appearance of $u$ in it.\;
      Delete $P$ from $\embedding (\theta_i )$. \; 
      $y$ := $\min_{u'v'\in \{uv\} \cup P}\left( f^{\theta_i}_{u'v'} \right)$ \;
      $f^{\theta}_{u'v'}$ := $f^{\theta}_{u'v'} - y \,\, \forall u'v'\in \{uv\} \cup P$
     }
     \Else{
     	$z(u)$ := $\min\left(z(v),f^{\theta_i}_{uv}\right)$ \;
     }
     \If{$u \neq v$}{
     	 Prefix $u$ in $\embedding(\theta_i)$ \;
       $v$ := $u$ \;
       Jump to the second statement inside the {\bf for} loop \; 
      }
      \Else{
       $\embedding(\eta)$ := $u, \forall \eta \in \preedges{\theta_i} $ \;
      }
   }
   $x(\embedding)$ := $z(s_1)$ \tcp*{Flow extracted on $\embedding$}
   $\lambda'$ := $\lambda' + x(\embedding)$ \tcp*{Total flow extracted}
\tcp{Remove $x(\embedding)$ amount of flow from all the edges in $\embedding$.}
   $f^{\theta}_{u'v'}$ := $f^{\theta}_{u'v'} - x(\embedding)$ $\forall \theta \in \compedges$ and $\forall u'v' \in B(\theta)$ \; 
\tcp{Remove $x(\embedding)$ amount of flow from all the relevant self-loops.}
   $f^{\theta}_{v'v'}$ := $f^{\theta}_{v'v'} - x(\embedding)$ $\forall \theta \in \compedges$ and $ v' = \first{B(\theta)}$ \; 
}
\end{algorithm}

{\em Proof of correctness of Algorithm~\ref{schedulingalgo}:} 
The proof of the following statements ensures the correctness of the algorithm.
\begin{enumerate}
\item \label{item1} In the third line inside the {\it for} loop, there exists a $u \in 
\selfandneighbors{v}$ such that $f^{\theta}_{uv} > 0$.

\item If a cycle of redundant flow is found and removed in the first
{\it if} block inside the {\it for} loop, then the remaining flows still
satisfy the constraints in the LP with $\lambda$ replaced by $\lambda - \lambda'$.
\label{item2}

\item At the end of each iteration of the {\it while} loop, the remaining flows
still satisfy the constraints in the LP with $\lambda$ replaced by $\lambda - \lambda'$.
\label{item3}

\item The algorithm terminates in finite time. \label{item4}
\end{enumerate}

We now outline a proof of each of these statements. We prove
the statements 1)--3) for a certain iteration of the loops while assuming that
all the above claims are true in all the previous iterations of the {\it while}
and {\it for} loops.

{\em Proof of \ref{item1}:} The current values of the flows satisfy
all the constraints in the {\em Node-Arc LP} with $\lambda$ replaced by
$\lambda - \lambda'$. The algorithm ensures that in this step,
the total outgoing flow $\sum_{u\in \neighbors{v}}f^\theta_{vu} \geq z(v) > 0$.
So, by constraints \eqn{eq:general_conservation} and \eqn{eq:flow_termination}, the total of incoming and generated flows
$\sum_{u\in \selfandneighbors{v} } f^{\theta}_{uv} > 0$. Hence the statement
follows.

{\em Proof of \ref{item2}:} We will prove that a cyclic flow
on a cycle $v_1,v_2,\cdots, v_l, v_1$ satisfies all the constraints
in the {\em Node-Arc LP} with $\lambda = 0$. Then clearly after subtracting
this flow from the edges of the cycle, the remaining flows in the network
will still
satisfy the constraints with the same $\lambda$ as before.
For a cyclic flow of type $\theta$ of volume $y$, the flow values
are $f^{\theta}_{v_iv_{i+1}} = y$ for $i=1,2,\cdots, l-1$,
$f^{\theta}_{v_lv_{1}} = y$, and all other
flow values are equal to $0$. So, for any node, any nonzero incoming
flow is `compensated' by the same amount of outgoing flow of the same
type. All flow values in the self-loops are also $0$. So clearly these
flows satisfy the constraints in the LP with $\lambda = 0$. This completes
the proof.

{\em Proof of \ref{item3}:} Again, we will prove that the removed
$x(\embedding)$ amount of flows on the edges of an embedding and
on the self-loops themselves satisfy
the constraints in the LP with $\lambda = x(\embedding)$. Then the remaining
flows will also satisfy the constraints with $\lambda$ replaced by
$\lambda - x(\embedding)$. The subtracted flow values are
$f^\theta_{uv} = x(\embedding)$ for $uv \in \embedding (\theta )$,
$f^\theta_{uu} = x(\embedding)$ for
$u = \pathstart{\embedding (\theta )}$, and all other flow values $0$.
We can verify that these flows satisfy the constraints in the {\em Node-Arc
LP}.

{\em Proof of \ref{item4}:} The {\em Node-Arc LP} has $O(m |\compedges|)$ number of variables $f^\theta_{uv}$ and $f^\theta_{uu}$. 
Each deletion of flows through a cycle, or through an embedding, makes at least one of these variables 
 zero. Since the number of steps in each iteration is finite, the algorithm ends in finite time. \hfill{\rule{3mm}{3mm}}

It can be checked that the overall complexity of 
Algorithm~\ref{schedulingalgo} is $O(\kappa^2 m^2)$.

\subsection{Primal-dual algorithm and min-cost embedding}
\label{primal_dual_algo}

The {\em Node-Arc LP} and the subsequent algorithm to find an optimal
solution of the {\em Embedding-Edge LP} has polynomial-time complexity. 
For the multi-commodity flow problem, and for more general packing
problems, Garg and Konemann~\cite{garg} gave a faster primal-dual algorithm to find
an $\epsilon$-approximate solution. The algorithm uses a
hypothetical subroutine/oracle. For the multi-commodity flow problem,
the subroutine finds the shortest paths between the source-terminal
pairs.
We now give a similar fast algorithm
to find an $\epsilon$-approximate solution to the {\em Embedding-Edge
LP}.

We first provide the dual of the {\em Embedding-Edge LP}. The
dual has the variables $\Ell = (l(e))_{e\in \netedges}$ corresponding
to the capacity constraints in the primal.
The dual LP is given as follows.

\centerline{\rule{\columnwidth}{0.75pt}}

{\em Dual of Embedding-Edge LP:} Minimize
$D(\Ell)=\sum_{e\in \netedges} c(e) l(e)$ subject to 

1. Constraints corresponding to each $x(B)$ in primal: 
\begin{equation}
  \sum_{e \in \embedding} r_{\embedding}(e) l(e) \ge 1, \; \forall \embedding
\end{equation}

2. Non-negativity constraints:
\begin{equation}
  l(e) \ge 0, \; \forall e \in E
\end{equation}

\centerline{\rule{\columnwidth}{0.75pt}}

\noindent
We define the weight of an embedding $\embedding$ as
\begin{align}
& \weight{\Ell}{\embedding} = \sum_{e \in \embedding} r_{\embedding}(e) l(e). \nonumber
\end{align}

\noindent
It can be checked (similar to \cite{garg}) that the dual LP is equivalent
to finding $\min_{\Ell}  \frac{D(\Ell )}{\minweight{\Ell}}$,
where
\begin{align}
& \alpha_l = \min_{\embedding}\weight{l}{\embedding} \nonumber
\end{align}
is the cost of the minimum cost embedding for $\Ell$.

The {\em Embedding-Edge LP} is a fractional packing LP of the type considered
by Garg and Konemann~\cite{garg} and Plotkin \etal~\cite{plotkin}.
A polynomial time primal-dual algorithm was presented in \cite{garg} for
such LPs assuming the existence of an efficient oracle subroutine which
finds a `shortest path.' For a packing LP
$\max\left\{a^Tx|Ax \leq b, x \geq 0\right\}$ and its dual LP
$\min\left\{b^Ty|A^Ty \geq a, y\geq 0\right\}$, the shortest path is
defined as $\sum_i A(i,j)y(i)/a(j)$~\cite{garg}.
It is easy to see that for our LP, the `shortest path' corresponds to
the embedding with minimum weight,
 $\arg\min_{\embedding}\weight{\Ell}{\embedding}$.
Algorithm~\ref{PrimalDual} gives the instance of the primal-dual
algorithm for our problem.

\newcommand{\dist}[2]{\text{dist}_{#1}(#2)}

\begin{algorithm} \label{PrimalDual} \caption{Algorithm for finding  approximately optimal
    $x$ and $\lambda$} 
  \SetKwData{Left}{left}\SetKwData{This}{this}\SetKwData{Up}{up}
  \SetKwFunction{Union}{Union}\SetKwFunction{FindCompress}{FindCompress}
  \SetKwInOut{Input}{input}\SetKwInOut{Output}{output}
  
  \Input{Network graph $\net =(\netnodes,\netedges)$, capacities $c(e)$, set of source
    nodes $\netsources$, terminal node $t$, computation tree $\compgraph = (\compnodes, \compedges )$,
    the desired accuracy $\epsilon$ }

  \Output{Primal solution $\{x(\embedding), \embedding \in \setofembeddings\}$}
\BlankLine Initialize
  $l(e):= \delta/c(e)$, $\forall e \in \netedges, x(\embedding) :=0, \forall \embedding \in
  \mathcal{\setofembeddings}$ \;

  \While{$D(l) < 1$}{

    $\embedding^*$ := OptimalEmbedding($\Ell$) \tcp*{OptimalEmbedding($\Ell$) outputs $\arg\min_{B}\weight{\Ell}{B}$}
    $e^*$ := edge in $B^*$ with smallest $c(e)/r_{B^*}(e)$ \;
    $x(B^*)$ := $x(B^*) + c(e^*)/r_{B^*}(e^*)$ \;
    $l(e)$ := $l(e)( 1 + \epsilon
    \frac{c(e^*)/r_{B^*}(e^*)}{c(e)/r_{B^*}(e)} ), \; \forall e \in
    B^*$ \;  
  }
  $x(B)$ := $x(B)/ \log_{1+\epsilon} \frac{1+\epsilon}{\delta},
  \forall B$ \; 
\end{algorithm}

\newcommand{\prev}[2]{\sigma_{#1}\left(#2\right)}
\newcommand{\adist}[2]{\omega_{#1}\left(#2\right)}

We now describe, and then provide below, the subroutine OptimalEmbedding($\Ell$) which finds a minimum
weight embedding of $\compgraph$ on $\net$ with a given length function $L$.
For each edge $\theta_i$, starting from $\theta_1$, it finds a way to
compute $\theta_i$ at each network node at the minimum cost possible.
It keeps track of that minimum cost and also the `predecessor' node
from where it receives $\theta_i$. 
If $\theta_i$ is computed at that node itself then the predecessor
node is itself.
This is done for each $\theta_i$ by a technique similar to the
Dijkstra's algorithm. Computing $\theta_i$ for
$i\in \{1,2,\ldots, \numS\}$ at the minimum cost at a node $u$
is equivalent to finding the shortest path to $u$ from $s_i$.
We do this by using Dijkstra's algorithm. For any other $i$, the
node $u$ can either compute $\theta_i$ from $\preedges{\theta_i}$
or receive it from one of its neighbors. To take this into account, unlike
Dijkstra's algorithm, we initialize the cost of computing $\theta_i$
with the cost of computing $\preedges{\theta_i}$ at the same node.
With this initialization, the same principle of greedy node selection
and cost update as in Dijkstra's algorithm is used to find the optimal
way of obtaining $\theta_i$  at all the nodes. Finally, the optimal embedding is
obtained by backtracking the predecessors. Starting from $t$, we backtrack
using predecessors from which $\theta_{|\compedges|}$ is obtained, till we
hit a node whose predecessor is itself. This node is the start node of 
$\embedding (\theta_{|\compedges|})$ and the end node of 
$\embedding (\eta)$ for all $\eta \in \preedges{\theta_{|\compedges|}}$.
The complete embedding is obtained by continuing this process for
each $\theta_i$ in the reverse topological order.

\begin{procedure}\caption{OptimalEmbedding($\Ell$)}

  \SetKwData{Left}{left}\SetKwData{This}{this}\SetKwData{Up}{up}
  \SetKwFunction{Union}{Union}\SetKwFunction{FindCompress}{FindCompress}
  \SetKwInOut{Input}{input}\SetKwInOut{Output}{output}
  
  \Input{Network graph $\net =(\netnodes,\netedges)$, Length function $\Ell$, set of source
    nodes $\netsources$, terminal node $t$, computation tree $\compgraph = (\compnodes, \compedges )$.} 
  
  \Output{Embedding $B^*$ with minimum weight under $\Ell$} 
  \BlankLine
  
  \For{$i = 1$ \KwTo $|\compedges|$}{

    \If{$i \in \{1, 2, \ldots, \numS\}$}{
      $\adist{u}{\theta_i}$ := $\infty, \forall u \in V-\{s_i\}$ \;
      $\adist{s_i}{\theta_i}$ := $0$ and $\prev{s_i}{\theta_i}$ := $s_i$ \;
    }
    \Else{
      $\adist{u}{\theta_i}$ := $\sum_{\eta \in \preedges{\theta_i}} \adist{u}{\eta}, \forall u \in V$ \; 
      $\prev{u}{\theta_i}$ := $ u, \; \forall u \in V  $ \; 
    }
    $\Psi$ := $\emptyset$; $\barb{\Psi}$ := $V$ \;
    
    \While{$\abs{\Psi} < n$}{
      $v$ := $\arg \min_{u \in \barb{\Psi}} \adist{u}{\theta_i} $ \;
      $\Psi$ := $\Psi \cup \{v\}$ \;
      $\barb{\Psi}$ := $\Psi - \{v\}$ \;
      \ForEach{$u \in N(v)$}{
        \lIf{$\adist{v}{\theta_i} + l(uv) < \adist{u}{\theta_i}$}{
          $\adist{u}{\theta_i}$ := $\adist{v}{\theta_i} + l(uv)$ and $\prev{u}{\theta_i}$ := $v$ \;
        }
      }
    }
    
  }

  $B^*(\theta_{|\compedges|})$ := $t$ \;
  \For{$i = |\compedges|$ \KwTo $1$}{
  $u$ := $B^*(\theta_{i})$ \tcp*{valid, as $B^*(\theta_{i})$ consists of only a node at this step}
  	\While{$\prev{u}{\theta_i}\neq u $}{
        Prefix $\prev{u}{\theta_i}$ to $B^*(i)$ \;
        $u$ := $\prev{u}{\theta_i}$ \;
    }
    $B(\eta)$ := $u \; \forall \eta \in \preedges{\theta_i}$  \;
  }
\end{procedure}

{\bf Correctness of  OptimalEmbedding($\Ell$):} It is
sufficient to show that, during each phase $i$, the algorithm computes
optimal values for $\adist{u}{\theta_i}$ and $\prev{u}{\theta_i}$,
for each node $u$ in $\net$. We prove this by induction on the pair
$(i,|\Psi|)$ according to the lexicographic ordering.
For $i \in \{1,\ldots,\numS\}$ and for all $|\Psi|$,
this follows from the correctness of Dijkstra's algorithm. Now, assuming the optimality
of $\adist{u}{\theta_i}$ and $\prev{u}{\theta_i}$ till all iterations before
$(i,|\Psi|)$, we prove the statement for $(i,|\Psi|)$. Suppose $v$ is the 
element added to $\Psi$ in the current iteration. We consider two cases:

Case 1: $\Psi = \{v\}$: The cost of computing (and not receiving from another
node) $\theta_i$ at any node $u$ is $\sum_{\eta \in \preedges{\theta_i}} \adist{u}{\eta}$. The algorithm chooses $v$
which has the minimum $\sum_{\eta \in \preedges{\theta_i}} \adist{u}{\eta}$
among all nodes $u \in \netnodes$ and assigns 
$\adist{v}{\theta_i} = \sum_{\eta \in \preedges{\theta_i}} \adist{v}{\eta}$
and $\prev{v}{\theta_i} = v$. If these are not optimal, then it must
be more efficient for $v$ to receive $\theta_i$ which is computed
at some other node $u$. But that implies $\sum_{\eta \in \preedges{\theta_i}} \adist{u}{\eta} < \sum_{\eta \in \preedges{\theta_i}} \adist{v}{\eta}$, which
is a contradiction to the choice of $v$.

Case 2: $\{v\} \subsetneq \Psi$: Suppose there is a more efficient way of 
receiving $\theta_i$ at $v$ than from the node selected as 
$\prev{v}{\theta_i}$ and that is to compute $\theta_i$ at a node $u$
and receive it along a path $P_{u,v}$.
Let the corresponding cost be $\omega'_v(\theta_i)$.
First, if $u \in \Psi'$, then the present cost $\left(\leq \sum_{\eta
\in \preedges{\theta_i}} \adist{u}{\eta}\right)$ at $u$ is less than
the present value of $\adist{v}{\theta_i}$, which is a contradiction to the
choice of $v$. Thus $u\in \Psi$. Let $u'$ be the last node in $P_{u,v}$
from $\Psi$, and $v'$ be the first node in $P_{u,v}$ from $\Psi'$.
Then $\omega'_{v}(\theta_i) \geq \adist{u'}{\theta_i} + l(u'v')
\geq \adist{v'}{\theta_i} \geq \adist{v}{\theta_i}$ --- a contradiction.
Here the first inequality follows since $u'\in \Psi$.
The second inequality follows from the update rule followed during
the inclusion of $u'$ in $\Psi$. The last inequality follows from the
choice of $v$.

\vspace{4mm}
\noindent
{\it Complexity of OptimalEmbedding($\Ell$) and the primal-dual algorithm:}
Let us consider the first {\em for} loop in OptimalEmbedding($\Ell$).
Each iteration of this loop is the same as Dijkstra's algorithm
except for the initialization. Thus, the for loop, excluding the
initialization step, can be run in $O(m+n\log n)$ time using Fibonacci heap
implementation. The initialization step requires $O(n|\preedges{\theta_i}|)$
time for each iteration. The second {\em for} loop has $O(n\numS)$
complexity.  So the overall algorithm takes $O(\numS (m+n\log n))$ time.

The number of iterations in the primal-dual algorithm is of the order
$O(\epsilon^{-1}m\log_{1+\epsilon}(m))$. Thus the overall complexity
of the algorithm is $O\left(\epsilon^{-1}\numS m (m+n\log n) \log_{1+\epsilon}(m)\right)$.

\section{Extensions}
\label{sec:extensions}

\noindent
{\bf 1. Multiple trees for the same function:} It may be possible to compute a function in different sequences of
operations which are expressed by different computation trees.
For example, the `sum' function $f(X_1,X_2,X_3) = X_1+ X_2+X_3$ may be
computed by any of the computation sequences $\big((X_1+X_2)+X_3\big)$,
$\big(X_1+(X_2+X_3)\big)$, or $\big(X_2 + (X_1+X_3)\big)$.
In general, suppose multiple computation trees
$\compgraph_1, \compgraph_2, \ldots,
\compgraph_\numgraphs$ are given for computing the same function.
Let $\setofembeddings_i$ denote the set of all embeddings of $\compgraph_i$
for $i=1,2,\ldots, \numgraphs$.
Let $\setofembeddings = \cup_i \setofembeddings_i$ denote the set of all
embeddings. Under this definition of $\setofembeddings$, the {\em Embedding-Edge
LP} for this problem is the same as that for a single tree.
The new OptimalEmbedding($\Ell$) algorithm finds an optimal embedding for
each $\compgraph_i$ and chooses the one with minimum weight as the
optimal embedding in $\setofembeddings$.
This can be used in the same primal-dual algorithm to find an
$\epsilon$-approximate solution.

Some edges of different trees may represent an identical function of the
sources. For example, for the function $X_1+X_2+X_3 +X_4$, an
edge corresponding to the function $X_1+X_2$ is present in each of the
trees corresponding to $\Big(\big((X_1+X_2)+X_3\big)+X_4\Big)$,
$\big((X_1+X_2)+(X_3+X_4)\big)$, and $\Big(\big((X_1+X_2)+X_4\big)+X_3\Big)$.
For this reason, OptimalEmbedding($\Ell$) algorithm can be made more
efficient by running iterations for each function rather than each edge.
The initialization of $\adist{u}{\theta}$ changes 
correspondingly, to take into account all possible ways of computing that
function. Rest of the algorithm remains the same. 

The particular function $\Func (X_1,X_2,\ldots, X_\numS) =
X_1+X_2+\ldots +X_\numS$ is of special theoretical as well as
practical interest.
There are many, of the order of $\numS !$, sequences of additions of data
and corresponding trees to get this function.
With the above modification, our OptimalEmbedding($\Ell$) algorithm
has complexity exponential in $\numS$ and linear in $m$. As a result,
our primal-dual algorithm gives an $\epsilon$-approximate solution in
exponential complexity in $\numS$ and quadratic in $m$.
The problem is equivalent to the much investigated multicast
problem. For this problem, and consequently for the function `sum',
the oracle finds a minimum weight Steiner tree. This is well-known
to be NP-hard on $\numS$. Approximate (but not $\epsilon$-approximate
for any given $\epsilon$) polynomial complexity algorithms are known (see \cite{saad} and citations therein)
for finding a minimum weight Steiner tree. This can also be used
to find approximate solution to the multicast, and hence the `sum',
in polynomial complexity~\cite{saad}.

\vspace*{4mm}
\noindent
{\bf 2. Multiple functions and multiple terminals:} Suppose the network has multiple terminals $t_1,t_2,\ldots, t_\numT$
wanting functions $\Func_1(X^{(1)}),\Func_2(X^{(2)}),\ldots,
\Func_\numT(X^{(\numT)})$ respectively. Here $X^{(i)}$ is the data
generated by a set of sources $\netsources^{(i)}$. The sets
$\netsources^{(i)}; i=1,2,\ldots, \numT$ are assumed to be pairwise disjoint.
For each function $\Func_i$, a computation tree
$\compgraph_i$ is given. Let us consider the problem of communicating the
functions to the respective terminals at rates $\lambda_1,\lambda_2,\ldots,
\lambda_\numT$. The problem is to determine the achievable rate region
which is defined as the set of 
$\br = (\lambda_1,\lambda_2,\ldots, \lambda_\numT)$ for which a protocol
exists for transmission of the functions at these rates. This
region can be approximately found by solving either of the following problems.

(i) For any given non-negative weights $\alpha_1,\alpha_2,\ldots, \alpha_\numT$,
what is the maximum achievable weighted sum-rate $\sum_{i=1}^\numT
\alpha_i \lambda_i$?

For this problem, we consider embeddings of the computation trees
$\compgraph_i$ into the network for each terminal $t_i$. Let
$\setofembeddings_i$ denote the set of all embeddings of $\compgraph_i$.
Then the {\em Embedding-Edge LP} for this problem
is to maximize $\sum_{i=1}^{\numT} \alpha_i \sum_{B\in \setofembeddings_i}
x(B)$. The constraints are the same as before with $\setofembeddings$
defined by $\setofembeddings = \cup_i \setofembeddings_i$.
The weight of an embedding $\embedding \in \setofembeddings$ under
a weight function $\Ell$ is defined
as $\alpha_i \weight{\Ell}{B}$ if $\embedding \in \setofembeddings_i$.
The new OptimalEmbedding($\Ell$) algorithm finds an optimal embedding for
each $\compgraph_i$ and chooses the one with minimum weight.
This can be used in the same primal-dual algorithm to find an
$\epsilon$-approximate solution. It is also easy to obtain a
{\em Node-Arc LP} for this problem by minor modifications to that for
a single function computation at a single terminal.

(ii) For any non-negative demands $\alpha_1,\alpha_2,\ldots, \alpha_\numT$,
what is the maximum $\lambda $ for which the rates $\lambda \alpha_1,
\lambda \alpha_2, \ldots, \lambda \alpha_\numT$ are concurrently achievable?

Here, we define an embedding to be a tuple $\embedding
= (\embedding_1,\embedding_2,\ldots, \embedding_\numT)$, where $B_i\in
\setofembeddings_i$ is an embedding of the computation tree $\compgraph_i$.
The {\em Embedding-Edge LP} for this problem is the same as that for the single terminal problem 
with $ r_{\embedding}(e)$ defined as $r_{\embedding}(e) = \sum_{i=1}^{\numT} \alpha_i |\{\theta \in \compedges_i | e \mbox{ is a part of }
\embedding_i(\theta )\}|$ and $\setofembeddings = \setofembeddings_1 \times \setofembeddings_2 \times \ldots \times \setofembeddings_\numT$. 
The weight of an embedding $\embedding $ under
a weight function $\Ell$ is defined
as $\sum_{i=1}^{\numT} \alpha_i \weight{\Ell}{B_i}$.
The new OptimalEmbedding($\Ell$) algorithm finds an optimal embedding $B$
by separately finding optimal embeddings $B_i$ for
each $\compgraph_i$. This can be used in the same primal-dual algorithm to find an
$\epsilon$-approximate solution. Again, we can easily obtain a {\em Node-Arc
LP} by minor modification to that for a single function computation
at a single terminal.

3. {\bf Computing with a precision:} In practice, the source data may be real-valued, and communicating
such a data requires infinite capacity. In such applications, it
is common to require a quantized value of the function at the terminal
with a desired precision. This may, in turn, be achieved by quantizing
various data types with pre-decided precisions and thus different
data type may require different number of bits to represent them. Suppose
the data type denoted by $\theta$ is represented using $\precision{\theta}$
bits. Then the {\it Embedding-Edge LP} and its dual for this problem are
the same as before
except that the definition of $r_\embedding (e)$ is changed to
$r_\embedding (e) = \sum_{\theta \in \compedges :e \text{ is a part of
 } B(\theta )} \precision{\theta}$. In the {\it Node-Arc LP}, the
capacity constraints are changed to
\begin{eqnarray}
&&  \sum_{\theta \in \compedges} \left( f^{\theta}_{uv} + f^{\theta}_{vu}
\right) \precision{\theta} \le c(uv) , \; \forall uv \in \netedges . \nonumber
\end{eqnarray}
In the {\it OptimalEmbedding($\Ell$)} algorithm, $l(uv)$ is replaced
by $l(uv)\precision{\theta_i}$ inside
the {\bf foreach} loop.

4. {\bf Energy limitted sensors:} Suppose, instead of capacity constraints
on the links, each node $u \in \netnodes$ has a total energy $\energy (u)$.
Each transmission and reception of $\theta$ require the energy $\energy_{T,\theta}$
and $\energy_{R,\theta}$ respectively. Generation of
one symbol of $\theta$ or computation of one symbol of $\theta$ from
$\preedges{\theta}$ requires the energy $\energy_{C,\theta}$. The objective is to compute the function
at the terminal maximum number of times with the given total node energy
at each node.

For an embedding $\embedding$, if $\embedding (\theta) = v_1,v_2,\cdots,
v_l$, then $tr(\embedding (\theta)) = \{v_1,v_2,\cdots, v_{l-1}\}$
denotes the transmitting nodes, and $rx(\embedding (\theta)) = 
\{v_2,v_3,\cdots, v_{l}\}$ denotes the receiving nodes of $\theta$.
If $l=1$, then $tr (\embedding (\theta))
= rx (\embedding (\theta)) = \emptyset$. For $\embedding$,
the energy load on the node $u$ is given by
\begin{eqnarray}
& \energy_B (u) = 
& \sum_{\theta: \pathstart{\embedding (\theta)}= u} \energy_{C,\theta} 
+ \sum_{\theta: u \in tx (\embedding (\theta))} \energy_{T,\theta} 
 + \sum_{\theta: u \in rx (\embedding (\theta))} \energy_{R,\theta}.  \nonumber
\end{eqnarray}
The capacity constraint in the {\it Embedding-Edge LP} is replaced
by the energy constraint on the nodes
\begin{eqnarray}
\sum_{\embedding \in \setofembeddings} x(\embedding ) \energy_\embedding (u)
\leq \energy (u) \,\,\forall u \in \netnodes , \nonumber
\end{eqnarray}
where an empty sum is defined to be $0$.
The dual of the {\it Embedding-Edge LP} is:
Minimize
$D(\Ell)=\sum_{u\in \netnodes} E(u) l(u)$ subject to

1. Constraints corresponding to each $x(B)$ in primal:
\begin{equation}
  \sum_{u \in \embedding} E_{\embedding}(u) l(u) \ge 1, \; \forall \embedding
\end{equation}

2. Non-negativity constraints:
\begin{equation}
  l(u) \ge 0, \; \forall u \in V.
\end{equation}
The weight or cost of an embedding can be defined as
\begin{align}
& \weight{\Ell}{\embedding} = \sum_{u \in \embedding} E_{\embedding}(u) l(u). \nonumber
\end{align}
The {\it OptimalEmbedding($\Ell$)} is modified in the weight initialization
and weight update. The weight initialization is done as 
$\adist{s_i}{\theta_i}$ := $\energy_{C,\theta_i}$ for source data and
$\adist{u}{\theta_i}$ := $\energy_{C,\theta_i}+\sum_{\eta \in \preedges{\theta_i}} \adist{u}{\eta}$ for other data. The weight update at $u$ is now done
as $\adist{u}{\theta_i}$ := $\adist{v}{\theta_i} + \energy_{T,\theta_i}+\energy_{R,\theta_i}$ if $\adist{v}{\theta_i} + \energy_{T,\theta_i}+\energy_{R,\theta_i} < \adist{u}{\theta_i}$.
After suitable modification, the primal-dual algorithm with the modified {\it OptimalEmbedding(L)} algorithm
finds an $\epsilon$-approximate solution. 

In the {\it Node-Arc LP}, the capacity constraints are replaced
by energy constraints at the nodes:
\begin{align}
 \sum_{\theta \in \compedges} f^\theta_{uu} \energy_{C,\theta}
 + \sum_{\theta \in \compedges}\sum_{v \in \neighbors{u}} (f^\theta_{uv} \energy_{T,\theta} + f^\theta_{vu} \energy_{R,\theta}) & 
 \leq \energy (u) \,\,\,\,\, \forall u \in \netnodes . \nonumber
\end{align}

\section{Discussion and conclusion}
\label{sec:discuss}

In this paper, we have laid the foundations for network
flow techniques for distributed function computation. Though
we have obtained results for computation trees, we believe
that much of our techniques can be extended to larger classes
of functions, for instance, fast Fourier transform (FFT),
that can be represented by more general graphical structures
like directed acyclic graphs and hypergraphs where each edge
or hyper-edge represents a distinct function of the sources. 
The sum function discussed in Sec.~\ref{sec:extensions}
is one such function representable by a hypergraph.

\begin{figure}[ht]
\centering
\subfigure[The butterfly network. Each edge has capacity 1 bit/use]{
\includegraphics[width=1.5in]{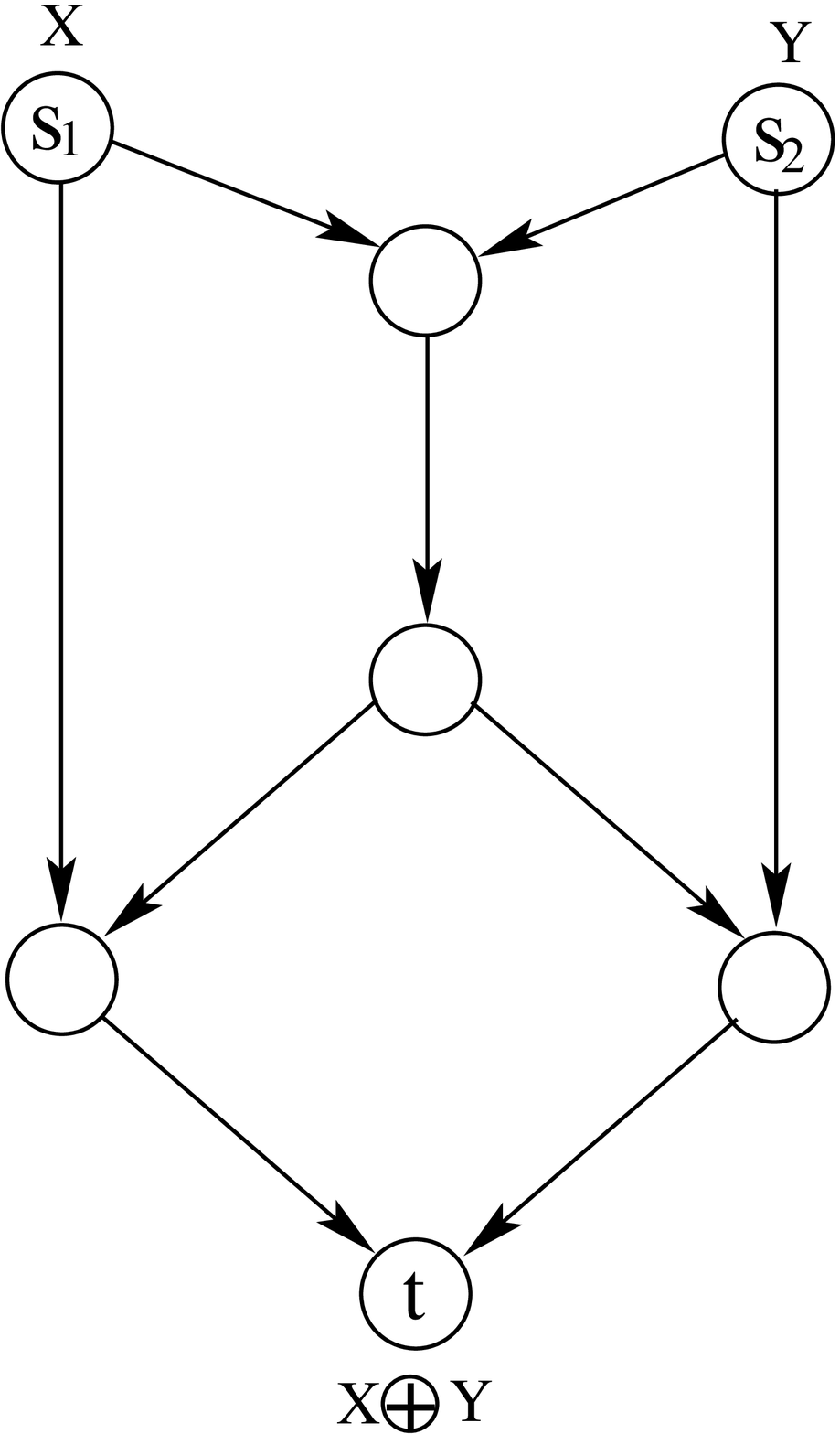}
\label{fig:butterfly1}
}
\hspace*{0in}
\subfigure[A rate-2 solution using cross-realization coding ]{
\includegraphics[width=1.4in]{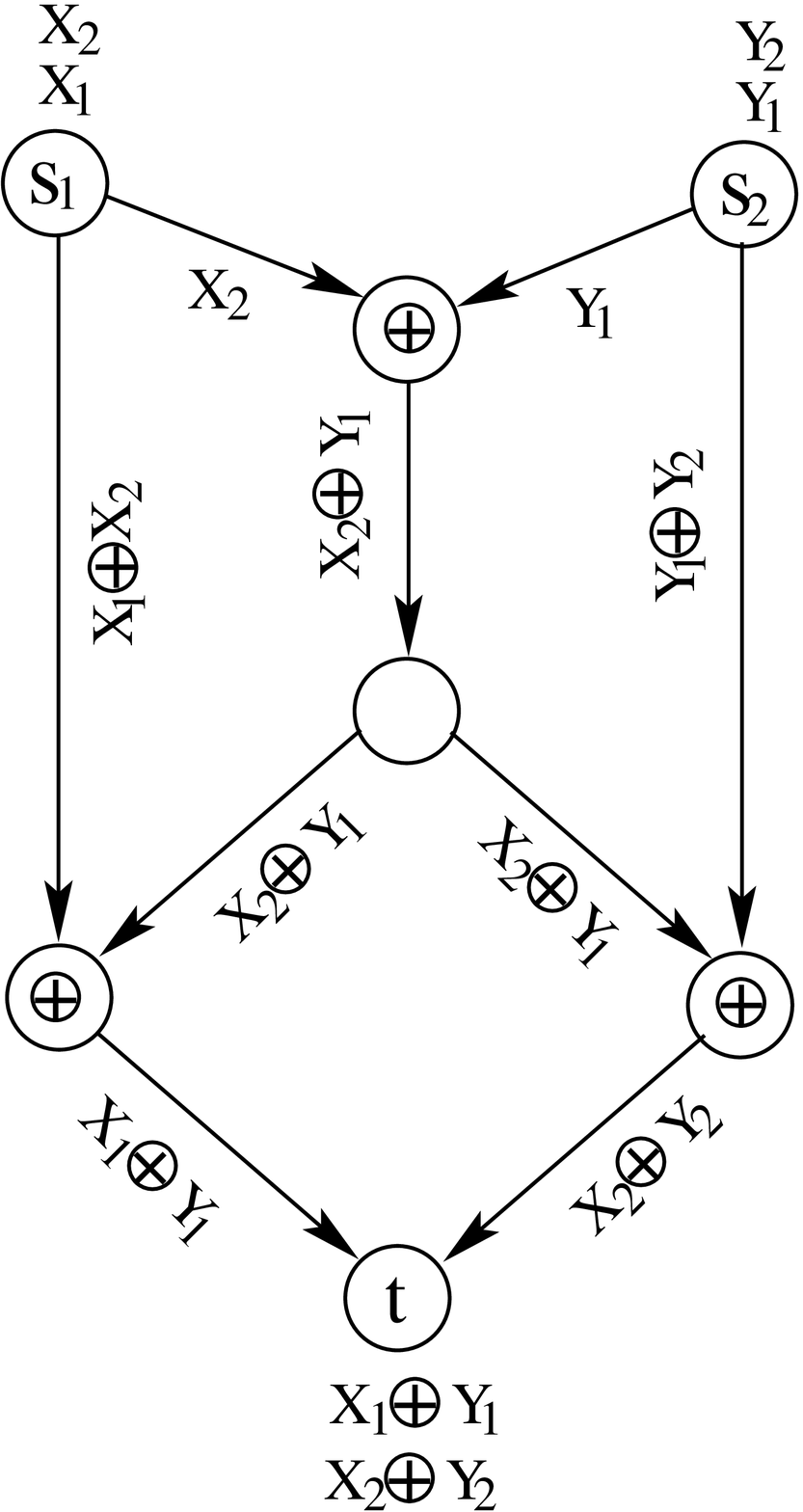}
\label{fig:butterfly2}
}
\caption[]{The butterfly network with XOR target function $\Func (X,Y) = X \oplus Y$}
\label{fig:butterfly}
\end{figure}

Our computation framework does not allow block coding, i.e., coding across
different realizations of the data. Such coding has been used
in the information theory and network coding literature.
Block coding can, in general, offer better computation rate.
For example, consider the directed butterfly network as shown
in Fig.~\ref{fig:butterfly} with two binary source nodes (with source processes
denoted by $X$ and $Y$) and a terminal node with a XOR target
function $\Func (X,Y) = X \oplus Y$. It can be checked that the maximum
rate achievable by routing-like schemes, i.e., without using inter-realization
coding, is $1.5$. On the other hand, the scheme shown in
Fig.~\ref{fig:butterfly2} using inter-realization coding achieves a
rate of $2$. However, for more general functions, finding the optimal
rate and designing optimal coding schemes is a difficult problem under
this framework. Further, for undirected multicast networks, it is known
that the inter-realization coding can achieve a rate strictly less
than twice the rate achieved by routing~\cite{lili}. We expect that
similar results will hold for function computation over undirected
networks. 

Altogether, we believe that results in this paper opens many new
avenues for further research.

\section{Acknowledgement}\label{section:acknowledgement}
The authors would like to thank A.~Diwan for fruitful discussions.
This work was supported in part by Bharti Centre for Communication at
IIT Bombay and a project from the Department of Science and Technology
(DST), India.

\bibliography{ref}
\bibliographystyle{IEEE}

\appendices
\section{The protocol}

 We now outline a communication and computation protocol designed to receive the
function at the terminal at a rate that is greater than
$\sum_{\embedding \in \setofembeddings} x(\embedding ) - \epsilon$ for
any given solution of the {\em Embedding-Edge LP}. First,
the flow values $x(\embedding )$ are rounded to lower rational numbers so
that the total flow $r$ is still greater than $\sum_{\embedding \in
\setofembeddings} x(\embedding ) - \epsilon$. With abuse of notation,
we use the same notation $x(B)$ to denote these rounded values of $x(B)$
in the rest of this subsection.
All these flows are then multiplied
by the least common multiple $N$ of the denominators of the flows 
$x(\embedding ); \embedding \in \setofembeddings$.
Let the resulting values be $n(\embedding ); \embedding \in \setofembeddings$.
Clearly $\sum_{\embedding \in \setofembeddings} n(B) = rN$.
Let us fix an order in the embeddings $B_1,B_2,\ldots, B_{|\setofembeddings|}$.
The protocol consists of computation at the nodes and communication across
the links in a block/frame of $N$ consecutive uses of the network.
In each frame, a link $e$ can carry upto a total of $Nc(e)$ symbols in both
directions. Our protocol will require sending integer number of symbols
in $N$ uses of $e$ in each direction. We assume that this is possible
as long as the total number of symbols transmitted in both directions is
at most $Nc(e)$. 
We assume that computation at nodes is done instantaneously, and a frame sent
across a link is available at the receiving node at the end of the frame.
The receiving node can forward the data on another edge in the next frame
or use it to compute something else for transmission in the next or
later frames.

In our protocol, 
the data stream generated at each source is divided into blocks of $rN$ symbols,
and the terminal computes $rN$ number of corresponding function values in
each frame. Out of the $rN$ computations, the first $n(\embedding_1)$
are carried out using the embedding $\embedding_1$, the next $n(\embedding_2)$
are carried out using the embedding $\embedding_2$, and so on. 
In each direction on each link, the transmissions corresponding to different embeddings
are ordered in the same order as the embeddings. Further, if
$uv$ is in $B(\theta_i)$ as well as $B(\theta_j)$ (assume $i<j$ without
loss of generality), then $uv$ carries the data for 
$(\embedding, \theta_i)$ first and then the data for $(\embedding, \theta_j)$.
Formally, in each frame and in each direction, a link $uv$ in $\net$ carries a subframe,
possibly empty, of data for each $(\embedding, \theta)$ pair, where
$\embedding \in \setofembeddings, \theta \in \compedges$. These subframes
are transmitted in the lexicographic order on $(\embedding, \theta)$.
Since the subframes for
different $(\embedding,\theta)$ may be available at $u$ with different delay,
these subframes will not correspond to the same frame of source data.
In the following, we explicitly describe the subframes carried by
$uv$ in the $k$-th frame.

Let $\by^k_{\embedding, \theta}$ denote the $n(B)$ symbols of data of
type $\theta$ corresponding to the $n(\embedding)$ symbols of
source data in the $k$-th frame corresponding to the embedding $\embedding$.
That is, $\by^k_{\embedding_1, \theta}$ denotes the $n(\embedding_1)$
symbols of data of type $\theta$ corresponding to the first $n(\embedding_1)$
symbols of source data in the $k$-th frame, $\by^k_{\embedding_2, \theta}$
denotes the $n(\embedding_2)$ symbols of data of type $\theta$
corresponding to the next $n(\embedding_2)$ symbols of source data in
the $k$-th frame, and so on.
In each frame, $uv$ carries a subframe of data for each
$(\embedding, \theta)$ pair. The subframe corresponding to
$(\embedding, \theta)$ is empty if 
$uv \not\in B(\theta)$. Formally,
\begin{align}
\by^k_{uv,\embedding, \theta}  & = \begin{cases} 
\by^k_{\embedding, \theta} \text{ if } uv \in B(\theta),\\
\emptyset \text{ otherwise.} \end{cases} \nonumber
\end{align}

This subframe corresponds to the $k$-th block of source data.
These subframes may
be available at $u$ with variable delay due to variable path lengths
from the sources along different embeddings. Let us define the depth
or delay $d(u,\embedding, \theta)$ as

\begin{align}
d(uv,\embedding, \theta) & = \begin{cases}
\infty \text{ if } uv \not\in \embedding (\theta ) \\
0 \text{ if } uv \in \embedding (\theta ), u = s_i, \theta = \theta_i \\
1+\max\{d(wu,\embedding, \eta) |\eta \in \preedges{\theta}, wu\in \embedding (\eta)\}\}  \\
\hspace*{3mm} \text{if } uv \in \embedding (\theta ), u = \pathstart{\embedding (\theta)}, (u,\theta) \neq (s_i, \theta_i) \\
d(wu,\embedding, \theta)+1 \text{ if } 
 (u,\theta) \neq (s_i, \theta_i), wu, uv \in \embedding (\theta ).
\end{cases}
\end{align}
So, the subframe $\by^k_{uv,\embedding, \theta}$, which has
$n(\embedding )$ symbols if $uv \in \embedding (\theta)$ and which corresponds
to the $k$-the frame of source data, will be
transmitted in the $\left(k+d(uv,\embedding, \theta)\right)$-th frame
on $uv$.
The infinite value for $uv \not\in \embedding (\theta )$ indicates
that the corresponding data does not flow through $uv$ from $u$ to $v$.

\begin{figure*}[!t]
\centering
\subfigure[A network to compute $\Func = X + Y.$]{
\includegraphics[width=.2\textwidth,height=0.25\textwidth]{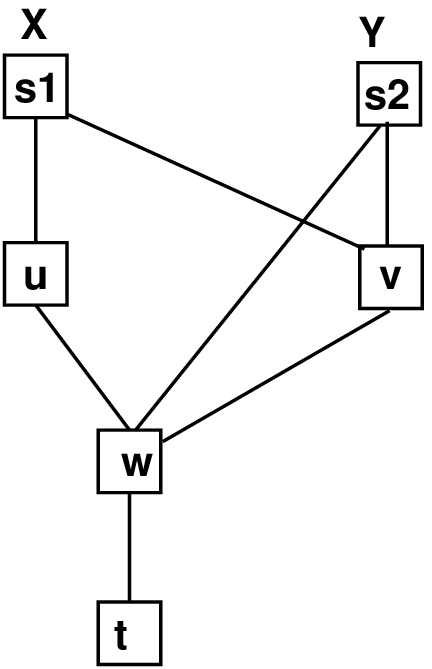}
\label{fig:net4}
}
\subfigure[A computation tree for $\Func.$]{
\includegraphics[width=.2\textwidth,height=0.25\textwidth]{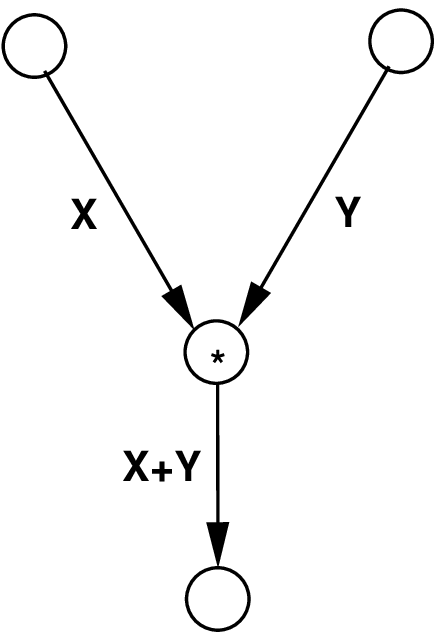}
\label{fig:net4tree}
}
\subfigure[A embedding $\embedding_1$.]{
\includegraphics[width=.2\textwidth,height=0.25\textwidth]{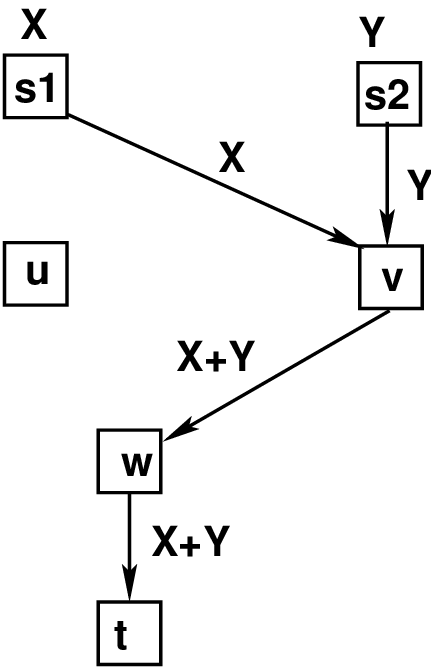}
\label{fig:net4embed1}
}
\subfigure[An embedding $\embedding_2$.]{
\includegraphics[width=.2\textwidth,height=0.25\textwidth]{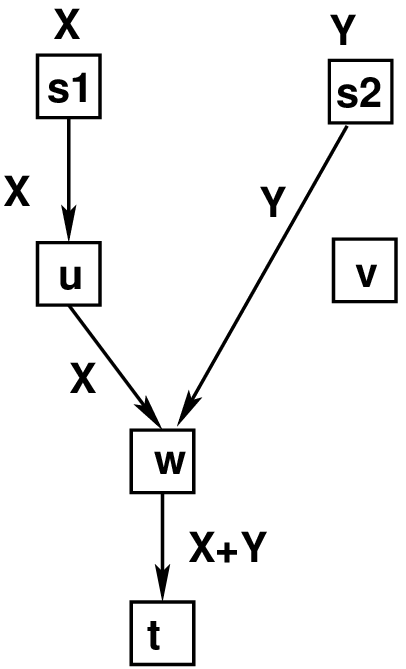}
\label{fig:net4embed2}
}
  \caption{A network, a computation tree and two embeddings}
\label{fig:example4}
\end{figure*}

\noindent
{\bf Example:} Consider the network and the computation tree shown in
Fig.~\ref{fig:example4}. The edges of the computation tree are labeled
by the functions they carry, that is, $X,Y,$ and $X+Y$.
For embedding $\embedding_1$, 
$d(s_1v, \embedding_1, X) = 0$, $d(s_2v, \embedding_1, Y) = 0$,
$d(vw, \embedding_1, X+Y) = 1$, $d(wt, \embedding_1, X+Y) = 2$,
and all other delay values are $\infty$.
For embedding $\embedding_2$,
$d(s_1u, \embedding_2, X) = 0$, $d(s_2w, \embedding_2, Y) = 0$,
$d(uw, \embedding_2, X) = 1$, $d(wt, \embedding_2, X+Y) = 2$,
and all other delay values are $\infty$.

The data transmitted in the $k$-th frame from $u$ to $v$ on the link $uv$,
in order of transmission, is thus
$\by_{uv, \embedding_1, \theta_1}^{k-d(uv, \embedding_1, \theta_1)},
\by_{uv, \embedding_1, \theta_2}^{k-d(uv, \embedding_1, \theta_2)},
\ldots, 
\by_{uv, \embedding_1, \theta_{|\compedges|}}^{k-d(uv, \embedding_1, \theta_{|\compedges|})},
\by_{uv, \embedding_2, \theta_1}^{k-d(uv, \embedding_2, \theta_1)},$
$\by_{uv, \embedding_2, \theta_2}^{k-d(uv, \embedding_2, \theta_2)},
\ldots,
\by_{uv, \embedding_2, \theta_{|\compedges|}}^{k-d(uv, \embedding_2, \theta_{|\compedges|})},
\ldots, 
\by_{uv, \embedding_{|\setofembeddings|}, \theta_1}^{k-d(uv, \embedding_{|\setofembeddings|}, \theta_1)},$
$\by_{uv, \embedding_{|\setofembeddings|}, \theta_2}^{k-d(uv, \embedding_{|\setofembeddings|}, \theta_2)},
\ldots,
\by_{uv, \embedding_{|\setofembeddings|}, \theta_{|\compedges|}}^{k-d(uv, \embedding_{|\setofembeddings|}, \theta_{|\compedges|})}$. It is easy to see that
the required flow of function values will be computed on each embedding
by this protocol. If the communication starts
with the frame number $0$ and ends with the $K$-th frame of source data,
then the subframes are empty for $k < d(uv, \embedding_i, \theta_j) $ and for
$ k > K + d(uv, \embedding_i, \theta_j) $.
In particular, a subframe $\by_{uv, \embedding_i, \theta_{j}}^{k-d(uv, \embedding_i,
\theta_{j})}$ is empty if $uv \not\in \embedding_i (\theta_j)$. 

\noindent
{\bf Example:} In the above example, suppose a solution of the
{\it Embedding-Edge LP} is $x(\embedding_1) = 1$ and $x(\embedding_2) = 0.5$.
Then $N = 2$, and $n(\embedding_1) = 2, n(\embedding_2) = 1$. Each
data stream is divided into frames of $3$ symbols, out of which
the first 2 symbols flow over $\embedding_1$ and the last symbol
flows over $\embedding_2$. In the
$k$-th frame, the link $uw$ carries only one non-empty subframe
for $\embedding_2$ containing one `$X$' symbol. That subframe 
$\by_{uw, \embedding_2, X}^{k-1}$ corresponds to the last symbol
of the $(k-1)$-th frame of data. The link $wt$ carries 
one subframe of two `$X+Y$' symbols for $\embedding_1$ and another
subframe of one `$X+Y$' symbol for $\embedding_2$. These subframes
$\by_{wt, \embedding_1, X+Y}^{k-2}$,$\by_{wt, \embedding_2, X+Y}^{k-2}$ 
correspond to the first two symbols of the $(k-2)$-th data frame and 
the last symbol of the $(k-2)$-th data frame respectively.

To implement the protocol, any node $u$ needs to know $N$, $n(\embedding)$
for all embeddings with non-zero $n(\embedding)$, and 
$d(uv, B, \theta )$ and $d(vu, B, \theta )$ for all such embeddings $B$,
$\theta \in \compedges, v \in \neighbors{u}$. The values of
$d(uv, B, \theta )$ can be found in $O(nb|\compedges|)$ time, where
$b$ is the number of embeddings for which $n(\embedding ) > 0$. In the following,
we give the sequence of actions taken by any node $u$.

1. The node maintains an
input queue for each $(\embedding, \theta)$ pair for which 
$d(vu, \embedding, \theta) < \infty$ for some $v\in \neighbors{u}$.

2. For the $k$-th frame received from $v$ on the link $vu$, the node
$u$ knows the `composition', i.e., how many symbols for which
$(\embedding, \theta)$ pair are received on that frame and in what
order. This is because the frame contains a non-empty subframe corresponding
to $(\embedding, \theta)$ if and only if $d(vu, \embedding, \theta) \leq k$.
Such a non-empty frame contains exactly $n(\embedding )$ symbols. 
The transmission of all the non-empty frames is ordered in the lexicographic
ordering of $(\embedding, \theta)$. For any received frame on any
link, $u$ puts each received subframe in its respective input
queue. If $u$ is a source, it also takes the $rN$ generated symbols
and creates the subframes of lengths $n(\embedding)$ for all
the relevant embeddings. Those are also placed in respective queues.

3. After queueing all the received and generated data in
the $k$-th frame, $u$ prepares the data
to be transmitted on each link $uv$ in the next, that is $(k+1)$-th,
frame of $N$ transmissions. The non-empty subframes for this transmitted frame 
are those for which $d(uv, \embedding, \theta) \leq k+1 $.
If there is an input queue for $(\embedding, \theta)$, i.e.,
if such a data subframe is received at $u$, then this subframe of
$n(\embedding)$ symbols is taken from the respective input queue.
Otherwise, this subframe is generated from the subframes from
the queues for $(\embedding, \eta); \eta \in \preedges{\theta}$.
If such a queue for $(\embedding, \eta)$ contains multiple
subframes of $n(\embedding)$ symbols, then the oldest of them is taken.
For instance, in our example (Fig.~\ref{fig:example4}), for
constructing the subframe $\by_{wt, \embedding_2, X+Y}^{k}$ 
at $w$ for the $k$-th frame, $w$ takes a subframe from
its input queue $(\embedding_2, X)$ and a subframe from the
input queue $(\embedding_2, Y)$ and adds them. At this time,
in the first queue, there is only one subframe $\by_{uw,\embedding_2,X}^{k-2}$
which is used now. But in the second queue, there are two subframes
$\by_{vw,\embedding_2,Y}^{k-1}$ and $\by_{vw,\embedding_2,Y}^{k-2}$
available, out of which the older subframe 
$\by_{vw,\embedding_2,Y}^{k-2}$ is used.

\end{document}